\documentclass[12pt]{article}

\usepackage{amsmath, amssymb}
\usepackage{graphicx}
\usepackage{titlesec}
\usepackage{pdflscape}
\usepackage{pdfpages}
\usepackage[round, authoryear]{natbib}
\newcommand{\blind}{0}

\begin{document}

\def\spacingset#1{\renewcommand{\baselinestretch}%
{#1}\small\normalsize} \spacingset{1}

\if0\blind
{
  \title{\bf Approximations for STERGMs Based on Cross-Sectional Data}
  \author{Chad Klumb\thanks{
    This work was supported by the National Institutes of Health under Grant R01-AI138783.  Partial support for this research came from a Eunice Kennedy Shriver National Institute of Child Health and Human Development research infrastructure Grant, P2C HD042828, to the Center for Studies in Demography and Ecology at the University of Washington.}\hspace{.2cm}\\
    Center for Studies in Demography and Ecology, University of Washington,\\
    Martina Morris \\
    Center for Studies in Demography and Ecology, University of Washington, \\
    Departments of Sociology and Statistics, University of Washington, \\
    Steven M. Goodreau \\
    Center for Studies in Demography and Ecology, University of Washington, \\
    Department of Anthropology, University of Washington, \\
    and
    Samuel M. Jenness \\
    Department of Epidemiology, Emory University}
  \maketitle
} \fi

\if1\blind
{
  \bigskip
  \bigskip
  \bigskip
  \begin{center}
    {\LARGE\bf Approximations for STERGMs Based on Cross-Sectional Data}
\end{center}
  \medskip
} \fi

\bigskip
\begin{abstract}
Temporal exponential-family random graph models (TERGMs) are a flexible class of network models for the dynamics of tie formation and dissolution.  In practice, separable TERGMs (STERGMs) are the subclass most often used, as these permit estimation from inexpensive cross-sectional study designs, and benefit from approximations designed to reduce the computational burden \citep{Carnegie}.  Improving the approximations are the focus of this paper.  We extend the work of \cite{Carnegie}, which addressed the problem of constructing a STERGM with two specific equilibrium properties: a cross-sectional distribution defined by a given exponential-family random graph model (ERGM), and tie durations defined by given constant hazards of dissolution.  We start with Carnegie et al.'s observation that the exact result is tractable in the dyad-independent case, and then show that taking the sparse limit of the exact result leads to a different approximation than the one they presented.  We show that the new approximation outperforms theirs for sparse, dyad-independent models, and that for dyad-dependent models the errors tend to increase with the level of dependence for both approximations.  We then extend the theoretical results of Carnegie et al. to the dyad-dependent case, proving that both the old and new approximations are asymptotically exact as the STERGM time step size goes to zero, for arbitrary dyad-dependent terms and some dyad-dependent constraints.  We also show that the continuous-time limit of the discrete-time approximations has exactly the combination of cross-sectional and durational equilibrium behavior that we seek.
\end{abstract}

\spacingset{1.5} 

\noindent%
\vfill

\newpage
\section{Introduction}
\label{sec:introduction}

One of the most general methodological frameworks for the statistical modeling of networks is the class of exponential-family random graph models (ERGMs).  These models specify probability distributions on networks as a function of the cross-sectional network statistics \citep{ergm}, and are commonly used as static network models.  Temporal exponential-family random graph models (TERGMs) are a discrete-time generalization of ERGMs that model transitions between network states in terms of (possibly time-dependent) network statistics  \citep{HX}.  Separable TERGMs (STERGMs) are a subclass of TERGMs in which the dynamics of edge formation and dissolution can be separated within each time step \citep{KH}.  For a review of these models and other terminology used in this introduction, see Section \ref{sec:tnc}.

While the statistical theory for TERGMs has been well developed \citep{HX, KH}, the practical estimation of these models remains a formidable computational challenge.  When complete longitudinal (panel) network data are available, the relatively efficient conditional MLE method can be used \citep{KH}.   In application contexts where the network data must be actively collected, rather than passively scraped, the cost to collect panel data can be prohibitive, so it is common to have only a single observation of a cross-sectional network state (or an estimate of its relevant statistics from an egocentric sample), with retrospective information on timing of edges.  A gradient descent algorithm based on the equilibrium generalized method of methods estimator (EGMME) has been developed to estimate the TERGM from this type of data \citep{K}, but the algorithm has not proven to be generally successful, owing to the computational burden of simulating TERGMs and estimating the gradient with finite differences, and the lack of a good method for initializing coefficients in the general case.

To improve the efficiency of estimation in the EGMME context, \cite{Carnegie} proposed a method for approximating the coefficients of a class of STERGMs by adjusting the coefficients of an ERGM for (possibly heterogeneous) edge durations with constant dissolution hazards. The resulting estimates can be used either as initial values for the EGMME STERGM algorithm, or directly (when the conditions favor sufficiently good approximations).  The goal of this method is to produce a STERGM with a cross-sectional equilibrium distribution that matches the given ERGM distribution, and equilibrium edge durations that match those defined by the constant dissolution hazards.  It gains efficiency by leveraging the relative ease of estimating ERGMs \citep{Hummel}, specifically in the context of egocentrically sampled data \citep{ego}, and has come to be called the ``edges dissolution approximation'' (EDA).

Implementing the EDA as presented in \cite{Carnegie} involves the following steps.  First, an ERGM containing the cross-sectional terms of interest is fit to data for a single cross-sectional network (possibly specified by target statistics).  The STERGM formation model terms are the same as those in the ERGM, and the STERGM dissolution model terms are a dyad-independent submodel of the ERGM.  Next, the dissolution coefficients are estimated directly from the target (mean) edge durations using a constant hazard model, equating the dissolution probability to the reciprocal of the target duration.\footnote{More precisely, the temporal data are the age of currently active edges.  Mean edge age is observed for active edges, while mean edge duration is unobserved, because it is censored for edges that are currently active.  Under a constant hazard model for dissolution, however, these two means are equal because the right-censoring exactly offsets the length bias, so estimating the mean edge age provides an estimate of the mean edge duration.}  Finally, the formation coefficients are obtained by taking the ERGM coefficients and subtracting off the corresponding dissolution coefficients.  This approximation is analogous to one commonly used in biostatistics and epidemiology when representing the relationship between disease prevalence, incidence and duration \citep{Alho}.

STERGMs have been used successfully in a wide range of applied research contexts \citep{Jasny_2019, OBrien, Park, Stivala}.  The primary application area for the EDA to date has been epidemic modeling \citep{JG, J}, where STERGMs are used as a principled, data-driven modeling framework for simulation studies.  The EDA makes it possible to estimate complex dynamic network models from cost-effective cross-sectional public health survey data on contact networks.  The fitted models are used to simulate dynamic networks that vary stochastically around the specified observed patterns, a transmission process is overlaid and the epidemic outcomes are tracked as they evolve under different scenarios.  EDA STERGMs are the default method for the EpiModel software package, which has been used in dozens of scientific publications \citep{epimodel}.

Despite its success, the EDA has some limitations.  The limitations are often linked to the type of model:  dyad-independent (where the edge state of a given dyad is not allowed to depend on the edge states of other dyads) or dyad-dependent (where it is allowed).  As noted in \cite{Carnegie}, the approximation error tends to rise as edge duration and network density fall for both model types, though in the dyad-dependent case the error can be nontrivial even when edge durations are not particularly short.  The authors showed that it is possible to reduce this error in the dyad-independent case by decreasing the size of the time step, but this introduces a potentially large computational burden.  Error reduction in the dyad-dependent case was not addressed theoretically.  The authors also assumed that the dissolution model was a subset of the ERGM model.  This condition has been viewed as an unwelcome restriction in applied work, where the factors influencing edge formation and edge duration may be different.

We address several of these shortcomings in the present paper.  In Section \ref{sec:digeneral} we derive a new form of the EDA that is adapted to sparse models.  We show in Section \ref{sec:dierror} that this new approximation outperforms the old approximation for sparse, dyad-independent models, in the sense that it reduces the error in cross-sectional edge probability.  In Section \ref{sec:sim}, we explore the behavior of both approximations for various dyad-dependent models, noting that while the new approximation is nearly exact for sparse models near dyad independence, there are regimes where the errors are large for both approximations.  In Section \ref{sec:dd}, we turn to theoretical results for dyad-dependent models.  We prove that both the old and new approximations are asymptotically exact as the size of the STERGM time step goes to zero, for arbitrary dyad-dependent terms and some dyad-dependent constraints.  We also show that the continuous-time limit of the discrete-time approximations has exactly the combination of cross-sectional and durational equilibrium behavior that we seek.  Our definition of EDA STERGMs in Section \ref{sec:tnc} allows us to clarify that the dissolution model need not be a subset of the ERGM model, and our results apply to this more general implementation of the EDA.  We discuss the implications of these findings and their place in the broader literature on dynamic network models in Section \ref{sec:discussion}.

\section{Materials and Methods}
\label{sec:methods}

\subsection{Software}
\label{sec:sw}

We will use a mix of formal derivations and simulations in this paper.  Simulations use R \citep{R} and the statnet suite of packages \citep{statnet}.  Model terms used in this paper are as implemented in the \texttt{ergm} package \citep{ergmpkg} for specifying the statistics of the network.

\subsection{Terminology, Notation, and Conventions}
\label{sec:tnc}

We begin by reviewing some basic network-theoretic terminology.  A network consists of a set of nodes and a set of edges.  It may be directed (in which case an edge is an ordered pair of nodes) or undirected (in which case an edge is an unordered pair of nodes).  A dyad is a potential edge; we will also identify a dyad with an ordered pair of nodes in the directed case and with an unordered pair of nodes in the undirected case.

Throughout this paper, the node set will be fixed but arbitrary, so that we can identify a network state with its edge set.  We will thus permit ourselves the use of standard set-theoretic notation and operations on network states, including union ($\cup$), intersection ($\cap$), difference ($\setminus$), symmetric difference ($\Delta$), membership ($\in$), containment ($\subseteq$), cardinality ($|\cdot|$), and empty set ($\varnothing$).  In particular, if $d$ is a dyad and $u$ is a network state, the notation $d \in u$ means $d$ is an edge in $u$.

In general, we may impose constraints on the network state, allowing only certain edge sets of a given directedness.  We will call a network state \emph{valid} if its edge set satisfies the constraints.  We will use $x,y,$ and $z$ to refer to valid network states; these symbols will always denote \emph{valid} network states, whether or not we use the term valid in referring to them.  A dyad $d$ is called \emph{free} if its edge state is not (unconditionally) fixed by the constraints, i.e., if there are valid network states $x$ and $y$ with $d \in x$ and $d \notin y$.  A dyad that is not free is called \emph{fixed}.  An edge is called \emph{free} (resp. \emph{fixed}) if its underlying dyad is free (resp. fixed).  A choice of constraints is called \emph{dyad-independent} if any free dyad can have its edge state toggled in any valid network state without violating the constraints.  Constraints that are not dyad-independent are called \emph{dyad-dependent}.

A mapping from the space of all network states of a given directedness to the real numbers is called a \emph{network statistic}.  We will use the term \emph{network statistics} to refer to a vector-valued function, each component of which is a network statistic.  (The case of a single component is allowed.)  We say that network statistics $g$ are \emph{dyad-independent} if for all network states $u,v$ (not necessarily valid) and all dyads $d$ we have $g(u \cup \{d\}) - g(u \setminus \{d\}) = g(v \cup \{d\}) - g(v \setminus \{d\})$.  Network statistics that are not dyad-independent are called \emph{dyad-dependent}.

An ERGM on a given node set is defined by a choice of directedness, constraints, network statistics $g$, and canonical coefficients $\theta$.  The network statistics and canonical coefficients combine to determine the cross-sectional distribution $\pi$ of the ERGM by the formula
\begin{equation} \label{eqERGM} \pi(x) = \frac{1}{C}\exp(\theta \cdot g(x)) \end{equation}
where $x$ is a valid network state and
$$ C = \sum_y \exp(\theta \cdot g(y)), $$
the sum being taken over all valid network states $y$.  An ERGM is called \emph{dyad-independent} if its constraints and network statistics are both dyad-independent; otherwise, it is called \emph{dyad-dependent}.

A STERGM on a given node set is defined by a choice of directedness, constraints, formation statistics $g^+$, dissolution statistics $g^-$, formation coefficients $\theta^+$, and dissolution coefficients $\theta^-$.  In general, the STERGM transition probabilities may depend arbitrarily on the history of network states, but in this paper we will make the following simplifying assumptions: we take the constraints to be cross-sectional, and we assume that $g^+$ and $g^-$ are given by cross-sectional statistics.  This means that if the network at time $t$ is $x$, then the probability that the network at time $t + 1$ is $y$ is given by
\begin{equation} \label{eqstergm} T_{xy} = \frac{1}{C_x} \cdot \frac{\exp(\theta^+ \cdot g^+(x \cup y))}{\exp(\theta^+ \cdot g^+(x))} \cdot \frac{\exp(\theta^- \cdot g^-(x \cap y))}{\exp(\theta^- \cdot g^-(x))} \end{equation}
where
$$ C_x = \sum_z \frac{\exp(\theta^+ \cdot g^+(x \cup z))}{\exp(\theta^+ \cdot g^+(x))} \cdot \frac{\exp(\theta^- \cdot g^-(x \cap z))}{\exp(\theta^- \cdot g^-(x))}. $$
These models are STERGMs in the sense of \cite{KH} except when the constraints are dyad-dependent, which breaks the separability of formation and dissolution within a time step.  We will use the term STERGM to refer to these models throughout this paper (understanding that in the case of dyad-dependent constraints, we should technically be using the more general term TERGM instead).  We say that a STERGM is \emph{dyad-independent} if its constraints, formation statistics, and dissolution statistics are all dyad-independent; otherwise, we say that the STERGM is \emph{dyad-dependent}.

An EDA STERGM is defined by a choice of ERGM, positive integer $L$, durational targets $\vec D \in [1,\infty)^L$, and mapping $\phi$ from dyads to $\{1,\dots,L\}$.  Given a network state $u$ and a value of $k \in \{1,\dots,L\}$, we let $u_k = \{d \in u : \phi(d) = k\}$.  We define network statistics $h$ mapping into $\mathbb{R}^L$ such that $h_k(u) = |u_k|$ for all network states $u$ and all $k \in \{1,\dots,L\}$.  Letting $g$ denote the network statistics of the ERGM and $\theta$ the canonical coefficients of the ERGM, we define the EDA STERGM (in the form of \cite{Carnegie}) as follows: its directedness and constraints match those of the ERGM; its formation statistics are $(g, h)$; its dissolution statistics are $h$; its formation coefficients are $(\theta, - \log(D_1 - 1), \dots, -\log(D_L - 1))$; its dissolution coefficients are $(\log(D_1 - 1), \dots, \log(D_L - 1))$.  Letting $\pi$ denote the ERGM distribution, this means that we can write \eqref{eqstergm} as
\begin{equation} \label{eqEDA} T_{xy} = \frac{1}{C_x} \cdot \frac{\pi(x \cup y)}{\pi(x)} \cdot \prod_k \frac{1}{(D_k - 1)^{|(x \Delta y)_k|}} \end{equation}
where
$$ C_x = \sum_z \frac{\pi(x \cup z)}{\pi(x)} \cdot \prod_k \frac{1}{(D_k - 1)^{|(x \Delta z)_k|}} $$
and $k$ ranges over $\{1,\dots,L\}$.  In general there are other choices of statistics $h$ and coefficients for the EDA STERGM that give rise to the same transition probabilities \eqref{eqEDA}; we have chosen the above for simplicity of description.  Note that even if $x$ and $y$ are valid, the network $x \cup y$ need not be valid (if the constraints are dyad-dependent).  We will assume that $\pi$ has been extended (by the formula \eqref{eqERGM}) to be defined on $x \cup y$ whenever $x$ and $y$ are valid, but that it is normalized with respect to the space of valid networks only.

In the context of EDA STERGMs, we say that a free dyad $d$ (or an edge on that dyad) is \emph{of type $k$} if $\phi(d) = k$.  The interpretation is that $D_k$ is the corresponding durational target.  We say that \emph{$y$ equals $x$ plus one edge of type $k$} if $x \subseteq y$ and $|y \setminus x| = |(y \setminus x)_k| = 1$.  We say \emph{$y$ equals $x$ minus one edge of type $k$} if $x$ equals $y$ plus one edge of type $k$.

We say (informally) that a network is \emph{sparse} if it has far fewer than the maximum number of edges given its directedness and the size of its node set, and that a model is \emph{sparse} if it tends to produce sparse networks.  For the precise results of Section \ref{sec:di}, edge probabilities of 1/3 or less are sufficient.

The arguments in this paper apply both to undirected models and to directed models.  Loops may be either allowed or forbidden, regarding the prohibition of loops as a particular dyad-independent constraint.  Our results also apply to bipartite models, regarding the bipartite condition as a particular dyad-independent constraint.  (A network is \emph{bipartite} if its vertex set can be partitioned into two subsets such that each edge in the network has one endpoint in each subset, and a model is \emph{bipartite} if its vertex set can be partitioned into two subsets such that the model constraints prohibit edges with both endpoints in the same subset.)

Where notation is specific to a section, we address it in that section.

\section{Results}
\label{sec:results}

\subsection{The Dyad-Independent Case}
\label{sec:di}

For the dyad-independent case we derive exact results for the formation coefficients following \cite{Carnegie}, define a new approximation for sparse models, derive the errors in both the old and new approximations, and derive the regime in which the new approximation performs better than the old approximation.

\subsubsection{Notation}
\label{sec:dinotation}

Suppose we are given a dyad-independent ERGM with statistics $g$ and coefficients $\theta$.  The linear predictor of a free dyad $d$ is defined to be $\eta_d = \theta \cdot (g(\{d\}) - g(\varnothing))$, which can be interpreted as the log odds of $d$ being an edge in a random network drawn from the ERGM distribution $\pi$.  In other words, if $p_d$ denotes the probability that free dyad $d$ is an edge under $\pi$, then $\eta_d = \text{logit}(p_d)$.  Since the ERGM is dyad-independent, its behavior is fully specified by its linear predictors for free dyads and its edge states for fixed dyads.  In particular,
$$ \pi(x) = \frac{\prod\limits_{\text{free}\, d \in x} \exp(\eta_d)}{\sum\limits_y \prod\limits_{\text{free}\, d \in y} \exp(\eta_d)}. $$

Likewise, the behavior of a dyad-independent STERGM of the type \eqref{eqstergm} is fully specified by its formation and dissolution linear predictors for free dyads, and its edge states for fixed dyads.  The formation linear predictor for free dyad $d$ is defined to be $\eta_d^+ = \theta^+ \cdot (g^+(\{d\}) - g^+(\varnothing))$, and the dissolution linear predictor for free dyad $d$ is defined to be $\eta_d^- = \theta^- \cdot (g^-(\{d\}) - g^-(\varnothing))$.  We can then write
$$ T_{xy} = \frac{1}{C_x} \cdot \left(\prod_{d \in (x \cup y) \setminus x} \exp(\eta_d^+)\right) \cdot \left(\prod_{d \in x \setminus (x \cap y)} \exp(-\eta_d^-)\right) $$
where
$$ C_x = \sum_z \left[\left(\prod_{d \in (x \cup z) \setminus x} \exp(\eta_d^+)\right) \cdot \left(\prod_{d \in x \setminus (x \cap z)} \exp(-\eta_d^-)\right)\right]. $$
The formation linear predictor $\eta_d^+$ for free dyad $d$ can be interpreted as the log odds of $d$ being an edge at time $t + 1$ given that it is not an edge at time $t$.  The dissolution linear predictor $\eta_d^-$ for free dyad $d$ can be interpreted as the log odds of $d$ being a edge at time $t + 1$ given that it is an edge at time $t$.

We also define $q_d$ to be the ``formation probability'' of free dyad $d$ as given by the relation $\eta_d^+ = \text{logit}(q_d)$, with the conventions that $\text{logit}(0) = -\infty$ and $\text{logit}(1) = +\infty$.

In the remainder of this section, $d$ will be fixed but arbitrary, and we will drop the subscript $d$ on $\eta_d, p_d, \eta_d^+, \eta_d^-$, and $q_d$.  We denote by $D$ the durational target for the dyad under consideration.

\subsubsection{Overview of Results}
\label{sec:dioverview}

As we are dealing with dyad-independent ERGMs, we will be able to focus on one free dyad at a time.  Expressed in terms of linear predictors, the prescription set forth in \cite{Carnegie} is to take
\begin{equation} \label{eq1}
\begin{aligned}
\eta^+ & = \eta - \log(D - 1), \\
\eta^- & = \log(D - 1).
\end{aligned}
\end{equation}
For dyad-independent STERGMs of the type \eqref{eqstergm}, specifying the dissolution linear predictor $\eta^-$ is equivalent to specifying the mean duration $D$ via the relation $\eta^- = \text{logit}(1 - 1/D) = \log(D - 1)$, with the conventions $\text{logit}(0) = \log(0) = -\infty$.  Given this value of $\eta^-$, the choice of value for $\eta^+$ is supposed to render equilibrium edge probability approximately that of the original ERGM.

In Section \ref{sec:digeneral}, we show that both the target mean duration $D$ and ERGM edge probability $p$ can be matched exactly by the STERGM when and only when $\frac{p}{(1 - p)D} \leq 1$, i.e. $D - \exp(\eta) \geq 0$, and that the STERGM linear predictors accomplishing this exact matching are uniquely given by
\begin{equation} \label{eq2}
\begin{aligned}
\eta^+ & = \eta - \log(D - \exp(\eta)), \\
\eta^- & = \log(D - 1).
\end{aligned}
\end{equation}

We are particularly interested in the sparse limit, which corresponds to $\eta \ll 0$; since $D \geq 1$, we can then approximate $\eta - \log(D - \exp(\eta))$ as $\eta - \log(D)$, so that the prescription \eqref{eq2} becomes
\begin{equation} \label{eq3}
\begin{aligned}
\eta^+ & = \eta - \log(D), \\
\eta^- & = \log(D - 1).
\end{aligned}
\end{equation}

We will refer to \eqref{eq1} as the ``old'' EDA, to \eqref{eq2} as the ``exact'' EDA, and to \eqref{eq3} as the ``new'' EDA.  While the exact result \eqref{eq2} could be implemented in practice (say, via an operator term \citep{ergm4}), it would generally require more than the simple model and coefficient adjustments that can be used to implement \eqref{eq1} and \eqref{eq3}.  For models that are very sparse across all dyad types, the difference between \eqref{eq2} and \eqref{eq3} should be negligible.

\subsubsection{Derivation of the Exact Solution and the Sparse Limit}
\label{sec:digeneral}

Here, we derive the general result \eqref{eq2} and (thus) obtain the sparse limit \eqref{eq3}.  The derivation of \eqref{eq2} is similar to that given in \cite{Carnegie}.  We use notation as in Section \ref{sec:dinotation}, for a fixed but arbitrary free dyad.  We take $D$ finite and $q$ positive, so that the finite irreducible Markov chain on edge states of this dyad (across time steps) possesses a unique stationary distribution.  We additionally let $p^*$ denote the equilibrium (cross-sectional) edge probability of this dyad under the STERGM.

We first derive a general relationship between $p^*,q,$ and $D$.  In equilibrium, the probability to form an edge on this dyad must equal the probability to dissolve an edge on this dyad, so as to conserve the total probability that there is an edge on this dyad.  We start with an edge with probability $p^*$, and given that we start with an edge, we dissolve it with probability $1/D$.  We start with a non-edge with probability $1 - p^*$, and given that we start with a non-edge, we form an edge with probability $q$.  Thus $(1 - p^*)q = p^*/D$, or $q = \frac{p^*}{(1 - p^*)D}$.

Note that given any two of $p^*, q$, and $D$, the relationship $q = \frac{p^*}{(1 - p^*)D}$ uniquely determines the third.  Thus, if we want to have $p^* = p$ for a given value of $D$, then we must have $q = \frac{p}{(1 - p)D}$, and since $q$ is a probability, this requires $\frac{p}{(1 - p)D} \leq 1$.  Conversely, suppose that $\frac{p}{(1 - p)D} \leq 1$, and define $q = \frac{p}{(1 - p)D}$.  Since $p^*$ is uniquely determined by the relationship $q = \frac{p^*}{(1 - p^*)D}$, it follows that $p^* = p$.

Thus, we can achieve cross-sectional edge probability $p$ and mean edge duration $D$ under the STERGM iff $\frac{p}{(1 - p)D} \leq 1$, and the value of $q$ accomplishing this is uniquely given as $q = \frac{p}{(1 - p)D}$.  We know $\eta^+ = \text{logit}(q)$ and $\eta = \text{logit}(p)$; using these relations together with $q = \frac{p}{(1 - p)D}$ yields after simplification that $\eta^+ = \eta - \log(D - \exp(\eta))$, as in \eqref{eq2}.

We now consider the sparse limit.  We have $\eta^+ = \text{logit}(q)$ and $\eta = \text{logit}(p)$; in the sparse limit $p \ll 1$, we may approximate $\text{logit}(p)$ by $\log(p)$, $\frac{p}{(1 - p)D}$ by $p/D$, and $\text{logit}(q) = \text{logit}\left(\frac{p}{(1 - p)D}\right)$ by $\log(p/D)$; the statement $\log(p/D) = \log(p) - \log(D)$ is then the statement that $\eta^+ = \eta - \log(D)$ in the sparse limit, so we have derived \eqref{eq3}.

Note that this argument is equivalent to that commonly used in biostatistics and epidemiology when representing the relationship between disease prevalence, incidence and duration (our $p$, $q$ and $D$ respectively) \citep{Alho}.  The general form of that relation is $\frac{p}{1-p} = qD$, and the equivalent ``sparse limit'' is $p = qD$.

\subsubsection{Formal Derivation of the Approximation Errors}
\label{sec:dierror}

In this section we derive and compare the errors of the old and new approximations \eqref{eq1} and \eqref{eq3}, assuming that the model and constraints are dyad-independent.  We focus on a single free dyad, with notation as in Section \ref{sec:dinotation}.  By the assumption of dyad independence, the mean duration will be matched exactly, so the error will be entirely in the equilibrium edge probability.  The target value is $p$, the ERGM edge probability, and we let $p_{\text{old}}$ and $p_{\text{new}}$ denote the equilibrium edge probabilities in the STERGM using the old and new approximations, respectively.  We can determine $p_{\text{old}}$ and $p_{\text{new}}$ by equating the exact results \eqref{eq2} for $p_{\text{old}}$ and $p_{\text{new}}$ to the approximations \eqref{eq1} and \eqref{eq3} for $p$ (with the same value of $D$ throughout, since we know all of \eqref{eq1}-\eqref{eq3} yield $D$ as the mean duration), and then solving for what $p_{\text{old}}$ and $p_{\text{new}}$ must be in order for these equations to be satisfied.

In order to do the derivation only once, we let $\alpha$ be a parameter taking values in $\{0,1\}$ and define $p_\alpha$ by
$$ \text{logit}(p_\alpha) - \log(D - \exp(\text{logit}(p_\alpha))) = \text{logit}(p) - \log(D - \alpha) $$
so that $p_{\text{new}} = p_0$ and $p_{\text{old}} = p_1$.  Noting that
$$ \text{logit}(p_\alpha) - \log(D - \exp(\text{logit}(p_\alpha))) = -\log\left(\frac{D}{\exp(\text{logit}(p_\alpha))} - 1\right) $$
we obtain
$$ -\log\left(\frac{D}{\exp(\text{logit}(p_\alpha))} - 1\right) = \text{logit}(p) - \log(D - \alpha). $$
Solving for $p_\alpha$, we find
$$ p_\alpha = p \cdot \frac{D}{D + p + \alpha(p - 1)}. $$
Thus we have the relative error
$$ \frac{p_\alpha - p}{p} = \frac{D}{D + p + \alpha(p - 1)} - 1 = \frac{-p - \alpha(p - 1)}{D + p + \alpha(p - 1)}. $$
This means that
$$ \frac{p_{\text{old}} - p}{p} = \frac{-2p + 1}{D + 2p - 1} $$
and
$$ \frac{p_{\text{new}} - p}{p} = \frac{-p}{D + p}. $$

We would like to identify for what values of $p$ and $D$ we have
$$ |p_{\text{new}} - p| < |p_{\text{old}} - p| $$
or, equivalently,
$$ \left|\frac{p_{\text{new}} - p}{p}\right| < \left|\frac{p_{\text{old}} - p}{p}\right|. $$
From the formulas above, this condition is equivalent to
$$ pD + 2p^2 - p < |-2p + 1|(D + p). $$
If $p > 1/2$, this becomes
$$ pD + 2p^2 - p < (2p - 1)(D + p) = 2pD + 2p^2 - D - p $$
i.e.
$$ 0 < (p - 1)D $$
which has no solutions as $p < 1$ and $D \geq 1$.  If instead $p \leq 1/2$, the condition is
$$ pD + 2p^2 - p < (1 - 2p)(D + p) = D + p - 2pD - 2p^2, $$
i.e.
$$ 4p^2 - p(2 - 3D) - D < 0. $$
The quadratic equation
$$ 4p^2 - p(2 - 3D) - D = 0 $$
has roots
$$ \frac{2 - 3D \pm \sqrt{4 + 4D + 9D^2}}{8} $$
and $4p^2 - p(2 - 3D) - D < 0$ precisely when $p$ lies strictly between these roots, as the parabola opens upwards.  It is clear that the lower root
$$ \frac{2 - 3D - \sqrt{4 + 4D + 9D^2}}{8} $$
is negative, whereas $p$ is constrained to be positive.  The formula for the upper root
$$ \frac{2 - 3D + \sqrt{4 + 4D + 9D^2}}{8} $$
defines a positive, monotonically decreasing function of $D \in [1,\infty)$, with limiting values $\frac{\sqrt{17} - 1}{8}$ as $D \to 1^+$ and $\frac{1}{3}$ as $D \to +\infty$.

In other words, the error (relative or absolute) in the equilibrium edge probability is smaller with the new approximation precisely when $p < \frac{2 - 3D + \sqrt{4 + 4D + 9D^2}}{8}$, which in particular includes the range $p \leq \frac{1}{3}$ for any value of $D$.

\subsection{Behavior of Dyad-Independent Approximations for Dyad-Dependent Models}
\label{sec:sim}

In this section, we utilize simulations to examine how the old and new EDAs behave on models with some commonly used dyad-dependent terms.  Using the syntax from the \texttt{ergm} package, we will specify an ERGM with the \texttt{degree} term, that counts the number of nodes with a particular degree, and the \texttt{gwesp} term, that computes a geometrically weighted measure of the number of edgewise shared partners.  The \texttt{gwesp} term has been shown to possess certain desirable properties over a simple homogeneous measure of triangles \citep{gwesp1, gwesp2}, and can be interpreted as a triad bias: two nodes are more or less likely to have an edge between them based on the number of other nodes to which they both have an edge (i.e., the number of shared partners).

Note that the new EDA is implemented in the same manner as the old EDA described in Section \ref{sec:tnc}, except that the formation coefficients for the new EDA are $(\theta, -\log(D_1), \dots, -\log(D_L))$.

\subsubsection{Simulation Setup}
\label{sec:simsetup}

To exhibit the behavior of the old and new approximations near dyad independence, we performed a series of simulations of \texttt{edges + degree(1)} models on a 1000 node undirected network.  We used mean degrees of 0.7, 1.0, 1.3, and 2.0; the range 0.7-1.3 is taken from applied research on HIV, where STERGMs are used to summarize and simulate sexual transmission networks \citep{G, W}, and the value 2.0 matches that used for the simulation in \cite{Carnegie}.  We also used \texttt{degree(1)} targets of 200 to 600 in steps of 100, whose range includes the mean \texttt{degree(1)} value for an edges-only model of each mean degree used.  A dyad-independent model is therefore within the range of models we simulate for each mean degree.  We included durations of 15, 50, and 100, to exhibit how errors change with duration when all other variables are held fixed; these durations include the range used for the simulation in \cite{Carnegie}, and are on the order of durations in the applied work measured in weeks or months \citep{G, W}.

We also performed simulations for various \texttt{edges + degree(1) + degree(2) + gwesp(0.5, fixed = TRUE)} models on a 1000 node undirected network, analogous to those of the original paper, but using \texttt{gwesp} throughout (rather than a mix of \texttt{triangle} and \texttt{gwesp}).  We used a mean degree of 2.0, a \texttt{degree(1)} target of 200, a \texttt{degree(2)} target of 350, and \texttt{gwesp(0.5, fixed = TRUE)} targets ranging from 3 to 300 (considering that an isolated \texttt{triangle} counts as three \texttt{gwesp}).  We again included durations of 15, 50, and 100.

\subsubsection{Simulation Results}
\label{sec:simresults}

The results of the \texttt{edges + degree(1)} simulations described in Section \ref{sec:simsetup} are shown in Figure \ref{fig:edeg1}.  The expectation that the new approximation is nearly exact for sparse, dyad-independent models is then supported by the fact that at the dyad-independent value of \texttt{degree(1)} (dashed purple line), the error with the new approximation (the red line) is very close to zero (dashed green line), regardless of duration.  More generally, Figure \ref{fig:edeg1} may be understood as demonstrating that in a neighborhood of a sparse, dyad-independent model, the new approximation outperforms the old approximation.\footnote{Also included on these plots are the results of simulating the Markov chain R defined in Section \ref{sec:dd}.  We discuss these results in Section \ref{sec:discussdd}.}  The size of this neighborhood will vary from model to model, and in general we do not have any way of predicting how large it will be, or whether it will contain specific dyad-dependent models of interest.  Both the new and old approximations tend to become more accurate as duration increases, consistent with observations made in \cite{Carnegie}.

The results of the \texttt{edges + degree(1) + degree(2) + gwesp(0.5, fixed = TRUE)} simulations are shown in Figure \ref{fig:gwesp}.  We found that substantially increasing the number of proposals per time step resulted in different trends for the old approximation than those shown in \cite{Carnegie}, suggesting that the higher number of proposals is needed to allow equilibration of the Metropolis-Hastings Markov chain within each time step.  A further tenfold increase in proposals (beyond the number used for Figure \ref{fig:gwesp}) produced largely similar results, suggesting the number used for Figure \ref{fig:gwesp} was sufficient to capture the main trends.  The errors show mixed results, with the new approximation generally outperforming the old for \texttt{edges}, \texttt{degree(2)}, and \texttt{gwesp}, but underperforming the old for \texttt{degree(1)}.  This illustrates the point that for an arbitrary dyad-dependent model, there is no guarantee which approximation will be better, regardless of duration; the new approximation tends to do better than the old approximation near dyad independence, but if we stray sufficiently far from dyad-independent models, we cannot predict which approximation will be better, and results may be mixed even within a single model.

\subsubsection{Interpretation}
\label{sec:siminterpretation}
Intuitively, we may think of the EDA errors specific to dyad-dependent models as arising from the differences between the networks used to specify the ERGM and the STERGM:  ERGMs use the cross-sectional network to compute the model statistics, as seen in \eqref{eqERGM}; STERGMs use the union network to compute the formation model statistics and the intersection network to compute the dissolution model statistics, as seen in \eqref{eqstergm}.  For any dyad-independent network statistics $g$ and any transition $x \to y$ between valid network states, the cross-sectional change statistics $g(y) - g(x)$ are simply the sum of the union change statistics $g(x \cup y) - g(x)$ and the intersection change statistics $g(x \cap y) - g(x)$.  This is not true in the dyad-dependent case, where the cross-sectional change statistics for a multi-toggle transition are not easily expressed in terms of the union and intersection change statistics.  However, when the transition is single-toggle, for any network statistics $h$ we have $h(y) - h(x) = h(x \cup y) - h(x) + h(x \cap y) - h(x)$, even if $h$ is dyad-dependent.

This suggests that reducing the amount of change per time step may reduce the EDA errors, which is empirically supported by the simulations in Figures \ref{fig:edeg1} and \ref{fig:gwesp}.  We prove a general result along these lines in the next section.

\subsection{Theoretical Results for Dyad-Dependent Models}
\label{sec:dd}

Our goal in this section is to prove that both the old and new EDAs are asymptotically exact as the size of the STERGM time step goes to zero, for models with arbitrary dyad-dependent terms and some dyad-dependent constraints.  In order to do this, we introduce in Section \ref{sec:ddinf} a discrete-time process, denoted $R$, that is related to the continuous-time limit of the EDA STERGMs, and that has the desired cross-sectional and durational behavior not just asymptotically but at any sufficiently small time step.  In Section \ref{sec:ddTERGM}, we obtain the desired asymptotic results for the EDA by comparing the EDA STERGMs to $R$ as the time step size goes to zero.\footnote{The asymptotic cross-sectional exactness result can be generalized as follows.  Suppose $F$ is a map from non-negative numbers $t$ to transition probability matrices on some finite state space, such that $F(0)$ is the identity, $F(t)$ is one-sided differentiable at $t = 0$, and the one-sided derivative $F^\prime(0)$ has a one-dimensional left kernel.  Then the left kernel of $F^\prime(0)$ is spanned by a (unique) probability vector $\pi$, and given any $\epsilon > 0$ there exists a $\delta > 0$ such that $0 < t < \delta$ implies that any stationary distribution $\sigma$ of $F(t)$ satisfies $||\sigma - \pi|| < \epsilon$, where $||\cdot||$ denotes the Euclidean norm.  The proof of this more general result is analogous to the one presented here.  Related convergence results (e.g. for finite-dimensional distributions) have appeared in the literature \citep{fdd, fdd2}.}

\subsubsection{General Setup and Notation}
\label{sec:ddnotation}

Suppose we are given an ERGM with arbitrary terms and constraints.  To prove cross-sectional exactness results, we will need the following assumption on the constraints:
\begin{enumerate}
\item[(i)] given any valid network states $x$ and $y$, there is a sequence $z^{(1)},\dots,z^{(m)}$ of valid network states such that $z^{(1)} = x$, $z^{(m)} = y$, and $|z^{(i)} \Delta z^{(i + 1)}| = 1$ for all $i \in \{1,\dots,m-1\}$.
\end{enumerate}
To prove durational exactness results, we will need the following additional assumption on the constraints:
\begin{enumerate}
\item[(ii)] given any valid network state, any edge in that network that is not unconditionally fixed by the constraints (i.e., any free edge) can be toggled off without violating the constraints.
\end{enumerate}
We will consider duration only for free edges.

Suppose we are also given a vector of positive durations $\vec D_0$ of length $L$, and a mapping $\phi$ from dyads to $\{1,\dots,L\}$.  We define $\vec D = \lambda \vec D_0$, where $\lambda$ is a positive scalar whose value we will regard as a variable parameter.  The interpretation is that $\vec D_0$ is given in some specific units (say days, weeks, years, ...) and $\lambda^{-1}$ signifies the fraction of that time unit that is represented by a single STERGM time step, making $\vec D$ the vector of durations in units of the STERGM time step.  We will frequently use the asymptotic notation $\mathcal O$; the limit being taken is $\lambda \to +\infty$, and the implied bounds may be interpreted as holding componentwise when stated for entire vectors or matrices.

The notation $||\cdot||$ denotes the Euclidean norm for vectors.

\subsubsection{Leading Order Behavior of Discrete-Time EDAs at Small Time Step Sizes}
\label{sec:ddleading}
Let $T$ denote the discrete-time EDA STERGM transition probability matrix with duration $\vec D$ (and dyad mapping $\phi$).  We will use the old convention \eqref{eq1} in writing out the details below, but the conclusions apply just as well to the new convention \eqref{eq3}, whose transition probability matrix differs from that of the old by $\mathcal{O}(1/\lambda^2)$.

Recalling \eqref{eqEDA}, if $x$ and $y$ are any valid network states (including the possibility that $x = y$), we have that
$$ T_{xy} = \frac{1}{C_x}\frac{\pi(x \cup y)}{\pi(x)} \prod_k \frac{1}{(D_k - 1)^{|(x \Delta y)_k|}} $$
where
$$ C_x = \sum_z \frac{\pi(x \cup z)}{\pi(x)} \prod_k \frac{1}{(D_k - 1)^{|(x \Delta z)_k|}}. $$
Note that for any $k$, we have $\frac{1}{D_k - 1} = \mathcal O(1/\lambda)$, so that
$$ \frac{\pi(x \cup z)}{\pi(x)} \prod_k \frac{1}{(D_k - 1)^{|(x \Delta z)_k|}} = \mathcal O(1/\lambda^{|x \Delta z|}). $$
Thus
$$ C_x = 1 + \mathcal{O}(1/\lambda), $$
the 1 coming from the case $z = x$, and the $\mathcal{O}(1/\lambda)$ encompassing all cases $z \neq x$.  Consequently,
$$ T_{xy} = \mathcal O(1/\lambda^{|x \Delta y|}). $$
Thus, we may write
$$ T = I + \frac{A}{\lambda} + \mathcal{O}(1/\lambda^2) $$
where the matrix $A$ is given by
$$ A_{xy} = \begin{cases} 0 & \text{ if $|x \Delta y| \geq 2$} \\
\frac{\pi(x \cup y)}{\pi(x)} \cdot \frac{1}{D_{0,k}} & \text{ if $|x \Delta y| = |(x \Delta y)_k| = 1$} \\
- \sum_{z \neq x} A_{xz} & \text{ if $x = y$} \end{cases}. $$
Note that the transition rate matrix of the continuous-time limit of $T$ is proportional to $A$.

\subsubsection{Exactness of Infinitesimal Time EDAs}
\label{sec:ddinf}
We now define an ``infinitesimal time step EDA STERGM transition probability matrix'' $R$ for sufficiently large values of the parameter $\lambda$ (relating $\vec D$ and $\vec D_0$ in the manner described in Section \ref{sec:ddnotation}) by declaring $R = I + \frac{A}{\lambda}$, where $I$ is the identity.  For $\lambda$ sufficiently large, this $R$ is a well-defined transition probability matrix on the state space of valid networks.  Note that the transition probability matrix $R$ defines a discrete-time process; motivation for the name ``infinitesimal'' and for the transition probabilities defining $R$ may be found in Section \ref{sec:discussinterpretations}.

The transition probabilities for $R$ prohibit direct transitions between networks that differ in edge state on two or more dyads; in this sense, multiple, simultaneous changes are forbidden.  Note that if $y$ equals $x$ plus one edge of type $k$, then $R_{xy} = \frac{\pi(y)}{\pi(x)} \cdot \frac{1}{D_k}$ and $R_{yx} = \frac{1}{D_k}$.  It follows that $R$ satisfies detailed balance with respect to the ERGM distribution $\pi$.  Since $R$ is finite and (under the assumptions stated in Section \ref{sec:ddnotation} for cross-sectional exactness) irreducible, $\pi$ is the unique stationary distribution of $R$.

Observe that under the assumptions stated in Section \ref{sec:ddnotation} for durational exactness, given any valid network state $x$ and any free edge of type $k$ in $x$, the probability under $R$ to transition from $x$ to the state $y$ that equals $x$ minus the edge in question is exactly $\frac{1}{D_k}$.  Under $R$, there are no other (positive probability) transitions out of $x$ that involve dissolving the edge in question, so this immediately implies that the mean duration of any free edge of type $k$ is $D_k$ (with a geometric distribution) under $R$.

We have thus shown that $R$ has precisely the cross-sectional and durational behavior we desire of the EDA: it reproduces the ERGM's cross-sectional network distribution and the specified mean edge durations with geometric distributions, even though the ERGM may have dyad dependence.  For these reasons, it may be of interest to simulate $R$ itself.  This presents its own set of challenges, which we discuss briefly in Section \ref{sec:discussdd}.  In the present argument, we will use $R$ to approximate the discrete-time EDA STERGM transition probability matrix with duration $\vec D$ so as to show that this latter transition probability matrix has the desired durational and cross-sectional behavior asymptotically as $\lambda \to \infty$.

From Section \ref{sec:ddleading}, we have that the continuous-time limit of the EDA STERGMs has transition rate matrix proportional to $A$.  It follows that the continuous-time limit also has the desired properties of the dynamic network model, in the sense that its cross-sectional equilibrium distribution is $\pi$, and its durations are exponentially distributed with the desired means (up to an overall scaling of time).  We choose to work with $R$ rather than the continuous-time limit for two reasons:
\begin{enumerate}
\item[\textbullet] a conceptual preference for comparing one discrete-time process (a discrete-time EDA STERGM) to another discrete-time process ($R$), and 
\item[\textbullet] $R$ can be simulated using an ERGM Metropolis-Hastings algorithm (see Section \ref{sec:discussinterpretations}).
\end{enumerate}

\subsubsection{Asymptotic Exactness of Discrete Time EDAs}
\label{sec:ddTERGM}
We will use the notation and results from Sections \ref{sec:ddleading} and \ref{sec:ddinf}.  Recalling that
$$ R = I + \frac{A}{\lambda} $$
and
$$ T = I + \frac{A}{\lambda} + \mathcal{O}(1/\lambda^2) $$
we define a $\lambda$-dependent matrix $\delta$ by
$$ T = R + \delta $$
so that $\delta = \mathcal{O}(1/\lambda^2)$.

Since $\pi$ is the unique stationary distribution of $R$ for any $\lambda$ (sufficiently large that $R$ is well-defined as a transition probability matrix), the left nullspace of $A$ is precisely $\text{span}\,\{\pi\}$.  (If there were a (real) vector $\sigma \notin \text{span}\,\{\pi\}$ with $\sigma A = 0$, then as $\pi_x > 0$ for all valid $x$ we will have all entries of $\pi + \alpha \sigma$ positive for sufficiently small positive $\alpha$.  Renormalizing $\pi + \alpha \sigma$ to a probability vector produces a stationary distribution of $R$ that is distinct from $\pi$, contradicting uniqueness.)  The positive, continuous function $v \mapsto ||v A||$ on the compact set $\{v \in (\text{span}\,\{\pi\})^\bot : ||v|| = 1\}$ thus admits a positive lower bound $c > 0$.

Since the transition probability matrix $T$ is finite and irreducible for any $\lambda$, it possesses a unique $\lambda$-dependent stationary distribution which we choose to write as $\pi + \epsilon$ where $\epsilon$ is a $\lambda$-dependent perturbation to the ERGM distribution $\pi$.  We have
$$ \pi + \epsilon = (\pi + \epsilon)T = (\pi + \epsilon)(R + \delta) = \pi + \epsilon + \epsilon \frac{A}{\lambda} + (\pi + \epsilon) \delta. $$
Rearranging,
$$ \epsilon A = -\lambda(\pi + \epsilon)\delta = \mathcal{O}(1/\lambda) $$
since $\delta$ is $\mathcal{O}(1/\lambda^2)$ and $\pi + \epsilon$ is $\mathcal{O}(1)$ (being a probability vector).  We now write
$$ \epsilon = \beta \pi + \pi^\bot $$
where $\beta$ is a $\lambda$-dependent scalar and $\pi^\bot$ is a $\lambda$-dependent element of $(\text{span}\,\{\pi\})^\bot$.  By our observations above about the left nullspace of $A$, we have
$$ c||\pi^\bot|| \leq ||\pi^\bot A|| = ||\epsilon A|| = \mathcal{O}(1/\lambda) $$
thus showing that $||\pi^\bot|| = \mathcal{O}(1/\lambda)$ since $c > 0$.  Letting $\vec 1$ denote the vector of 1s, we have
$$ 1 = \vec 1 \cdot (\pi + \epsilon) = \vec 1 \cdot \pi + \vec 1 \cdot \epsilon = 1 + \vec 1 \cdot (\beta \pi + \pi^\bot) = 1 + \beta + \vec 1 \cdot \pi^\bot $$
thus showing
$$ \beta = -\vec 1 \cdot \pi^\bot = \mathcal{O}(1/\lambda). $$
Since $\epsilon = \beta \pi + \pi^\bot$ and both $\beta$ and $\pi^\bot$ are $\mathcal{O}(1/\lambda)$ (while $\pi$ is of course $\mathcal{O}(1)$), we have $\epsilon = \mathcal{O}(1/\lambda)$.  Since $\pi + \epsilon$ is the stationary distribution of $T$, this shows that the stationary distribution of $T$ converges to $\pi$ as $\lambda \to \infty$, proving asymptotic cross-sectional exactness for EDAs with dyad-dependent ERGM model terms (and some dyad-dependent constraints).

We still need to note that mean durations are asymptotically correct, under the required assumptions stated in Section \ref{sec:ddnotation}.  This is true in the relative sense: for any $0 < \epsilon < 1$ (having nothing to do with the $\epsilon$ above), there is a $\lambda_0(\epsilon)$ such that $\lambda > \lambda_0(\epsilon)$ implies all free edges of type $k$ have mean duration between $(1 - \epsilon)D_k$ and $(1 + \epsilon)D_k$ under $T$.  The reason is that free edges of type $k$ have dissolution probability exactly $\frac{1}{D_k}$ under $R$ and $T = R + \mathcal{O}(1/\lambda^2)$; taking $\lambda$ sufficiently large (depending on the given $\epsilon$), we can guarantee that the dissolution probability under $T$ of any free edge of type $k$ in any valid network state is between $\frac{1}{(1 + \epsilon)D_k}$ and $\frac{1}{(1 - \epsilon)D_k}$, thus guaranteeing its mean duration is between $(1 - \epsilon)D_k$ and $(1 + \epsilon)D_k$ (and its cumulative distribution function is sandwiched between those of a geometric distribution with mean $(1 + \epsilon)D_k$ and a geometric distribution with mean $(1 - \epsilon)D_k$).

\section{Discussion}
\label{sec:discussion}

We have presented a new form of the EDA that is tailored to sparse models, proven asymptotic exactness of both the old and new EDA for a broad class of dyad-dependent models, and shown that the continuous-time limit of the EDA STERGMs and its discrete-time relative $R$ have precisely the cross-sectional and durational properties we desire of the dynamic network model.  Here, we provide some further commentary on these results.

\subsection{When to Expect Better Results from the New EDA}
Generally speaking, we expect the greatest benefit of the new approximation \eqref{eq3} over the old approximation \eqref{eq1} to be for sparse models with weak dyad dependence and short duration.  While \eqref{eq3} may be better than \eqref{eq1} for some strongly dyad-dependent models, there does not seem to be a good way to tell when this is the case short of simulating both approximations and examining the errors.  The two approximations become equivalent in the long-duration limit, in the sense of the asymptotics presented in Section \ref{sec:dd}.

\subsection{Dissolution Model Need Not Be a Submodel of the ERGM}
In the presentation of \cite{Carnegie}, the dissolution model for the EDA STERGM was assumed to be a submodel of the original ERGM.  It should be clear from Section \ref{sec:tnc} that this restriction is unnecessary: given the ERGM, the durational targets may vary arbitrarily, free dyad by free dyad.  

The approach in Section \ref{sec:tnc} represents the most general form of the EDA: the dissolution model is specified with a statistic for each dyad type (as determined by the durational targets), and each such statistic is therefore also included in the EDA STERGM formation model.  In practice, dissolution models summarize the systematic patterns in edge dissolution using common dyad-independent terms, possibly depending on nodal or dyadic attributes.  The adjustment principle is the same in both cases, however:  the EDA STERGM formation model coefficients are obtained by subtracting the coefficients of the dissolution model (with or without the durational adjustment of +1, for the new and old EDA respectively) from the coefficients of the ERGM.  When a term appears in both the dissolution model and the ERGM, we subtract one coefficient from the other.  When a term only appears in the dissolution model, the dissolution model coefficient is subtracted from zero to calculate the corresponding STERGM formation coefficient.

To give a simple example of this approach, again using the syntax from the \texttt{ergm} package, consider an ERGM model specified with $\sim$\texttt{edges}, and durational targets that vary according to whether or not nodes match on \texttt{"sex"}, so the dissolution model can be taken to be $\sim$\texttt{edges + nodematch("sex")}.  By implication, the formation model for the EDA STERGM is then $\sim$\texttt{edges + nodematch("sex")}.  Letting $\theta$ denote the \texttt{edges} coefficient in the ERGM, $D_0$ the durational target for edges not matching on \texttt{"sex"}, and $D_1$ the durational target for edges matching on \texttt{"sex"}, the dissolution coefficients are $\log(D_0 - 1)$ for the \texttt{edges} term and $\log(D_1 - 1) - \log(D_0 - 1)$ for the \texttt{nodematch} term.  The formation coefficients are then approximated by $\theta - \log(D_0)$ for the \texttt{edges} term and $\log(D_0) - \log(D_1)$ for the \texttt{nodematch} term, using the new EDA.

\subsection{Interpretations of R}
\label{sec:discussinterpretations}
With notation as in Section \ref{sec:dd}, from our analysis of $R$ and the discrete-time EDA STERGM transition probability matrix $T$, we can see that if we expand $T$ in powers of the small parameter $\lambda^{-1}$, then $R$ corresponds exactly to keeping the 0th and 1st order terms in this expansion, dropping all higher order terms.  Thus, if we express the STERGM time step $dt$ as $dt \propto \lambda^{-1}$, then we obtain $R$ from $T$ by regarding $dt$ as a formal infinitesimal: $dt$ itself is not zero, but $(dt)^2 = 0$.  This is why we call $R$ an infinitesimal time STERGM, and also shows how $R$ is related to the continuous-time limit of $T$.

The literature on dynamic network modeling contains several continuous-time models with known general cross-sectional ERGM equilibria.  Examples include LERGMs \citep{lergm} and certain SAOMs \citep{saom}.  What distinguishes $R$ from these continuous-time models is that it reproduces both the cross-sectional ERGM statistics and the independently specified edge duration targets. This makes $R$ quite useful from an application perspective, when the goal is to reproduce a dynamic network with specific structure and edge dynamics that are either observed in data, or designed as an experiment.  Indeed, this was the original motivation for the STERGM EDA:  the purpose of the research was to explore, via simulation, the impact of variations in network structure and dynamics on the spread of infectious diseases, with models that, once estimated from empirical data, would reproduce the specified empirical patterns with fidelity.

There is also a relationship between $R$ and the Metropolis-Hastings chain for the ERGM, for particular choices of proposal.  More explicitly, if $E$ denotes the transition probability matrix for the ERGM Metropolis-Hastings chain with proposal $P$, then for any valid network states $x$ and $y$ with $P(y | x) > 0$, we have
$$ E_{xy} = \min\left(\frac{\pi(y)}{\pi(x)} \cdot \frac{P(x|y)}{P(y|x)},1\right) \cdot P(y | x). $$
Now, we assume that
\begin{enumerate}
\item[\textbullet] $P(y|x) = 0$ if $|x \Delta y| \geq 2$, and
\item[\textbullet] $P(y|x) = \frac{1}{D_k}$ when $y$ equals $x$ minus one edge of type $k$, and
\item[\textbullet] $P(y|x) \geq \frac{\pi(y)}{\pi(x)} \cdot \frac{1}{D_k}$ when $y$ equals $x$ plus one edge of type $k$.
\end{enumerate}
Then, when $y$ equals $x$ plus one edge of type $k$, we have
$$ E_{xy} = \frac{\pi(y)}{\pi(x)} \cdot P(x|y) = \frac{\pi(x \cup y)}{\pi(x)} \cdot \frac{1}{D_k} = R_{xy} $$
and when $y$ equals $x$ minus one edge of type $k$, we have
$$ E_{xy} = P(y|x) = \frac{\pi(x \cup y)}{\pi(x)} \cdot \frac{1}{D_k} = R_{xy}. $$
Since $E_{xy} = R_{xy} = 0$ when $|x \Delta y| \geq 2$, and both $R$ and $E$ are transition probability matrices, it follows that $R = E$.  We can thus identify one proposal in the ERGM Metropolis-Hastings chain with one time step in the infinitesimal time STERGM.  Note that it is always possible to satisfy the above conditions on the proposal $P$ whenever $\lambda$ is large enough that $R$ is well-defined as a transition probability matrix, by taking equality in the third bullet point and assigning any remaining proposal probability to the diagonal $P(x|x)$.

\subsection{Dyad-Dependence and Efficient Computation}
\label{sec:discussdd}
The approximations \eqref{eq1} and \eqref{eq3} were derived under the assumption of dyad independence.  While their performance may be adequate in the context of weak dyad dependence or long duration, they can both have substantial errors for short-duration, strongly dyad-dependent models, regardless of density.  The general way to improve their behavior while continuing to use discrete-time STERGMs is to shrink the size of the STERGM time step (i.e., to increase all durations by a single multiplicative factor).  The validity of this approach (first suggested in \cite{Carnegie}) is justified by the asymptotics in Section \ref{sec:dd}.

In principle, one could avoid these errors entirely by instead using the ``infinitesimal time STERGM'' $R$ defined in Section \ref{sec:ddinf}.  As proven there, $R$ has precisely the cross-sectional and durational behavior we are trying to achieve with the EDA, even for dyad-dependent models.  An example of the cross-sectional improvement of $R$ over the approximations \eqref{eq1} and \eqref{eq3} is shown in Figure \ref{fig:edeg1}.  To realize $R$ as a Metropolis-Hastings chain for the original ERGM, however, one must find a value of $\lambda$ such that $\max_x \sum_{y \neq x} R_{xy} \leq 1$, and implement a proposal satisfying the conditions in Section \ref{sec:discussinterpretations}.  As a practical matter, one may also wish to restrict the state space to avoid exceptionally unlikely networks, as this can allow a smaller value of $\lambda$ to be used.  Carrying these steps out without using a wastefully large value for $\lambda$ or a wastefully slow proposal can be nontrivial, especially for models with complex dyad dependence.

Finding an efficient algorithm is a general issue for dynamic network modeling, and existing software bears witness to this search.  For example, the choice of $\lambda$ and proposal for $R$ is analogous to the choice of burnin controls and proposal for simulating discrete-time STERGMs using the \texttt{tergm} package \citep{tergm}.  The four simulation burnin controls in the \texttt{tergm} package jointly affect how many proposals are used per discrete time step, similar to the way that $\lambda$ controls the number of proposals per unit real time when simulating $R$.  Just as $\lambda$ and the proposal cannot be chosen independently for $R$, the burnin controls and the proposal cannot be chosen independently for discrete-time STERGMs (without risking substantial errors in the simulations).

Broadly speaking, incorporating some aspects of the model being simulated into the proposal can improve efficiency for both $R$ and discrete-time STERGMs.  As noted above, when choosing $\lambda$ and the proposal for simulating $R$, strict quantitative conditions must be met in order to guarantee the validity of the simulation.  The simulation of discrete-time STERGMs does not come with such explicit conditions, however, failing to use enough proposals per discrete time step (given the choice of proposal) will generally produce inferior results.

Ultimately, whether $R$ is ``better'' than a discrete-time EDA STERGM for a given application will depend on the particulars of that application.  For ERGMs \citep{ergm4} and discrete-time EDA STERGMs, it has been possible to reduce computation time by orders of magnitude by using a judicious choice of model-aware proposal.  This suggests similar progress could be made in the future for $R$.

\subsection{Outlook}
Finally, we note that while we have focused exclusively on EDA STERGMs in this paper, the full class of TERGMs (and even STERGMs) is much more general.  While data limitations may preclude taking advantage of this generality in some cases, a more robust algorithm for estimating TERGMs in the EGMME context would be of great practical value.  For example, the ability to handle explicitly temporal effects and dyad-dependent dissolution hazards could improve network models as used in applied settings such as epidemic modeling, thus allowing better estimation of relevant transmission dynamics under various possible intervention scenarios.  Ultimately, a reliable method for estimating TERGMs from limited data may obviate the considerations of this paper.

\section{Acknowledgments}

We acknowledge Dave Hunter and Alina Kuvelkar for their review of the manuscript, Carter Butts for his review of the manuscript and discussions about continuous-time processes with ERGM equilibria, and the statnet development team for general support.

\section{Declaration of Interest Statement}
The authors report there are no competing interests to declare.

\bibliography{EDA}

\section{Figures}
\label{sec:figures}

\begin{landscape}
\includepdf[angle=90]{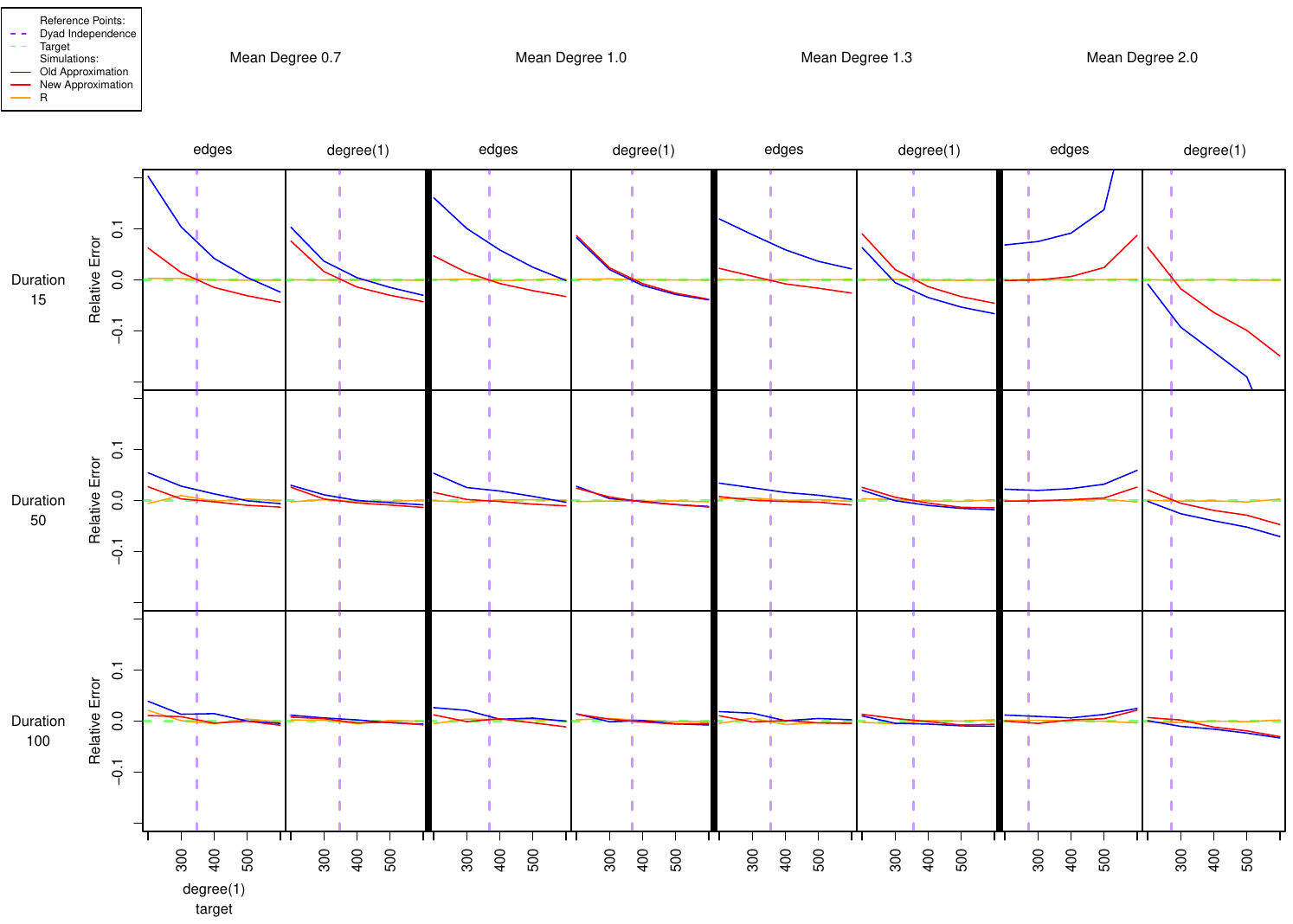}
\end{landscape}

\begin{landscape}
\includepdf[angle=90]{gwespplots.pdf}
\end{landscape}

\begin{figure}[h]
\caption{Relative errors for the \texttt{edges} and \texttt{degree(1)} statistics for the \texttt{$\sim$edges + degree(1)} model on a 1000 node undirected network with mean degree targets of 0.7-2.0 and a range of \texttt{degree(1)} targets and durations.}
\label{fig:edeg1}
\end{figure}

\begin{figure}[h]
 \caption{Relative errors for the \texttt{edges}, \texttt{degree(1)}, \texttt{degree(2)}, and \texttt{gwesp(0.5, fixed = TRUE)} statistics for the \texttt{$\sim$edges + degree(1) + degree(2) + gwesp(0.5, fixed = TRUE)} model on a 1000 node undirected network with mean degree 2.0, \texttt{degree(1)} target 200, \texttt{degree(2)} target 350, and a range of \texttt{gwesp(0.5, fixed = TRUE)} targets and durations.}
 \label{fig:gwesp}
\end{figure}

\end{document}